\documentclass[%
reprint,
superscriptaddress,
twocolumn,
amsmath,
amssymb,
aps,
prl
]{revtex4-1}

\usepackage{graphicx}  
\usepackage{dcolumn}  
\usepackage{bm}            
\usepackage{bbold}
\usepackage{amsmath}
\usepackage{xfrac}
\usepackage{xcolor}
\usepackage[colorlinks=true, citecolor=red]{hyperref}   
\usepackage{hhline}
\usepackage{diagbox}
\usepackage{siunitx}
\usepackage{enumitem}
\usepackage[normalem]{ulem}

\newcommand{\bmk}{\bm{k}}

\newcommand{\bmr}{\bm{r}}

\newcommand{\bmQ}{\bm{Q}}

\begin{document}
	
	\title{Collective Spin Excitations in Correlated Moir\'e Chern Ferromagnets}

	\author{Ming Xie}
	  	    \affiliation{Condensed Matter Theory Center, Department of Physics, 
			                University of Maryland, College Park, Maryland 20742, USA}
             \affiliation{Department of Physics, 
	                	The University of Texas at Dallas, Richardson, TX 75080, USA}
	\author{Sankar Das Sarma}
    	\affiliation{Condensed Matter Theory Center, Department of Physics, 
    		                University of Maryland, College Park, Maryland 20742, USA}
    	                
	\date{\today}
	
\begin{abstract}
Moiré-induced narrow electronic bands in transition metal dichalcogenide superlattices support many correlated quantum phases characterized by novel charge, flavor, and topological orders. Among these, magnetic ordering emerges as the most ubiquitous, often serving as the parent state for other correlated phases, including quantum anomalous Hall states, as well as chiral superconducting state. Because of electron-electron correlation, the stability of magnetic order is critically influenced by low-energy collective spin fluctuations, or magnon excitations. We investigate the nature of magnon excitations and their impact on the stability and transition temperature of the magnetic state at integer filling factor $\nu = -1$. We find that the magnon spectrum exhibits isolated low-energy bands whose topological character undergoes a transition upon tuning the interlayer displacement field. The magnon gap is found to depend sensitively on the topology of the magnetic ground state, resulting in an order-of-magnitude enhancement of the transition temperature $T_c$ in the quantum anomalous Hall phase compared to the topologically trivial correlated insulator. Our findings provide insight into the interplay between electron and magnon topology and suggest new routes for controlling magnetism and topology via moiré engineering.


\end{abstract}
	
\maketitle

\emph{Introduction}---\noindent
Twisted bilayer transition metal dichalcogenide (TMD) superlattices host narrow moir\'e bands with strong electronic correlations, giving rise to a diverse array of emergent quantum phenomena, including correlated insulating states, integer and fractional quantum anomalous Hall effects, and chiral superconductivity 
\cite{MakHubbard2020, WignerCrystal, Shan2020, Dean2020, LeRoy, StripePhase, MakMott2021, Pasupathy2021, Cui2021, ShanMIT,
QuantumCriticality, HeavyFermion, MakQAHE2021, Cai2023, Zeng2023, Park2023, Fan2023, Feldman2023, Tao2024, MakQSHMoTe2, MakQSHWSe2, XuFCI2026, CaiOpticalSwitching2025, MakVanHove2025, YankowitzTopology2025, GaoMoTe2FM2025, XuHigherFlatBand2025, LiSecondBand2025, YoungImaging, XuUniversalMagnetism2025, XiaSuperconductivity2025, GuoSuperconductivity2025, LiUnconventionalSC2025, DeanWSe2SC2025}. 
Many of these quantum phases exhibit hysteretic behavior under magnetic field sweeps and nonzero Hall resistivity at zero magnetic field as a consequence of spontaneous spin ordering \cite{ MakQAHE2021, Cai2023, Zeng2023, Park2023, Fan2023, Feldman2023, Tao2024, MakQSHMoTe2, MakQSHWSe2, XuFCI2026, CaiOpticalSwitching2025, MakVanHove2025, YankowitzTopology2025, GaoMoTe2FM2025, XuHigherFlatBand2025, LiSecondBand2025, XuUniversalMagnetism2025, LiUnconventionalSC2025}.
The magnetic behavior persists to temperatures higher than the transition temperature of these quantum phenomena, 
suggesting that magnetic ordering underlies these phases and 
is the energetically stable parental state from which they emerge.  The stability of magnetic order is therefore critical to the existence of these emergent phases, whose transition temperatures are ultimately bounded from above by that of the magnetic transition.

The stability of magnetic order in two-dimensional systems is strongly influenced by low-energy collective spin fluctuations. In particular, gapless magnon modes can inhibit the formation of true long-range order, allowing only quasi-long-range orders in systems with easy-plane anisotropy. By contrast, when magnon excitations are gapped, such as in systems with easy-axis anisotropy, long-range magnetic order persists at finite temperatures. In moiré superlattices formed from TMD layers, the combination of spin–valley locking and anisotropy in the valley degree of freedom gives rise to a range of magnetic ground states, including out-of-plane ferromagnetic phases (or valley-polarized states) and in-plane antiferromagnetic (or intervalley coherent) phases.
While mean-field theory has been instrumental in predicting candidate magnetic ground states, it fails to accurately describe systems where spin fluctuations beyond the mean-field approximation are significant. This is particularly true in two-dimensional TMD moir\'e systems where spin fluctuations are strongly enhanced due to the lack of intrinsic strong valley anisotropy. It is therefore crucial to understand the nature of the  spin fluctuations beyond mean field and their impact on magnetic ordering.

In this work, we report on a study of the collective spin excitations, or magnons, in twisted TMD homobilayers, 
focusing at one hole per moir\'e unit cell, corresponding to filling factor $\nu=-1$.
We use time-dependent Hartree-Fock calculations performed in both the full continuum model basis 
and projected moir\'e band basis to accurately assess the nature of the magnon excitations around the mean-field ground states. We find that the magnon spectrum features two low-energy bands isolated from the continuum of the spin-flip excitations. Upon varying the interlayer displacement field, the two magnon bands undergo a gap closing topological transition while the ground state remain in the QAH phase. 
The magnon gap drops drastically at the ground state phase transition
from the QAH insulator to the trivial correlated FM insulator. We calculate the Curie temperature by incorporating the effect of magnon thermal fluctuations  and find an order of magnitude decrease in the transition temperature across the QAHE transition.


\begin{figure}[t!]
	\centering
	\includegraphics[width=0.475\textwidth]{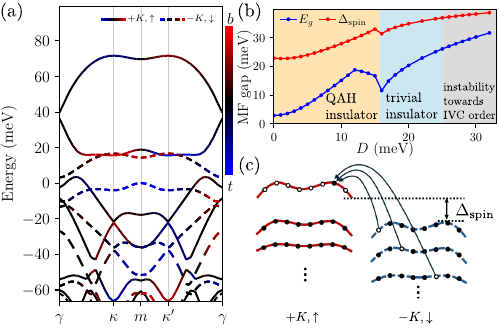}
	\caption{
		\label{MF_band}(a) Band structure of the FM ground state at twist angle $\theta = 3.5^\circ$ and zero interlayer displacement field obtained from the full self-consistent mean-field calculation. Solid (dashed) lines represet $+K(-K)$ valley and color indicates the wavefunction weight on the top (t) and bottom (b) layers. (b) Global charge gap ($E_{\rm g}$) and spin gap ($\Delta_{\rm spin}$) of the mean-field ground state as a function of displacement field. The orange and blue shading indicate regions with QAH and trivial FM insulating ground states, respectively, and the gray shading suggests the region where FM is unstable towards an IVC ground state. (c) Schematic diagram of spin-flip excitations from occupied spin down (-K) bands to the empty spin up (+K) bands.
	}
\end{figure}

\emph{Mean-field ground states}---\noindent
We first employ self-consistent Hartree-Fock calculations to obtain the mean-field ground states of twisted TMD homobilayers, using twisted MoTe$_2$ as a representative example. Our analysis focuses on filling factor $\nu = -1$, where magnetism is most prominent and mean-field theory is more reliable than at partial filling factors. The non-interacting electronic structure can be effectively captured by the moiré continuum Hamiltonian $\mathcal{H}_0$, which is constructed from the low-energy $\bm{k} \cdot \bm{p}$ model for the highest valence bands near the $+K$ and $-K$ valleys of the isolated monolayers \cite{WumoireTI}.
The non-interacting Hamiltonian takes the form
\begin{align}
&\mathcal{H}_0^{s} \negthinspace= \negthinspace
	\begin{pmatrix}
		H^0_{t}(\bm{k})\negthinspace+\negthinspace V_t(\bm{r})\negthinspace+\negthinspace D/2& T_{s}(\bm{r})\\
		T_{s}^\dagger(\bm{r})  & H^0_{b}(\bm{k})\negthinspace+\negthinspace V_b(\bm{r})\negthinspace-\negthinspace D/2
	\end{pmatrix}\negthinspace,\negthinspace \\
&\Delta_{t(b)}(\bm{r})= 2V\sum_{i=1,3,5} \cos(\bm{g}_{i}\cdot\bm{r}\mp\phi),\\
&T_s(\bm{r}) = t(1 + \omega^{s} e^{i\bm{g}_2\cdot\bm{r}} + \omega^{2s} e^{i\bm{g}_3\cdot\bm{r}}),
\label{moireHam}
\end{align}
where $s=\pm1$ for spin $\uparrow\negthickspace/\negthickspace\downarrow$, which is locked with the valley index $\{+K, -K\}$ due to spin-orbit interaction.
$H^0_{t(b)}(\bm{k})=-\hbar^2(\bm{k}-\bm{\kappa}(\bm{\kappa}'))^2/2m^*$ is the effective Hamiltonian 
for the top (bottom) layer with effective mass $m^*$, and
$D$ is the displacement field between the two layers.
$V$ and $\phi$ are the amplitude and phase of the intralayer moir\'e potential $V_l$.
$\bm{g}_1={4\pi}/{\sqrt{3}a_M}(1, 0)$ and $\bm{g}_i=(\hat{\mathcal{R}}_{\pi/3})^{i-1}\bm{g}_1$ 
are moir\'e reciprocal lattice vectors 
where $\hat{\mathcal{R}}_{\pi/3}$ is counter-clockwise rotation around $z$ axis by $\pi/3$.
$T_{s}(\bm{r})$ is the interlayer tunneling term for spin $s$ electrons 
with tunneling strength $t$ and phase factor $\omega=e^{i2\pi/3}$. 
The electron-electron interaction takes the form $V(\bm{q})=2\pi e^2/(\epsilon q) \tanh(qd)$ corresponding to the dual-gate screened Coulomb interaction. 
We perform self-consistent Hartree-Fock calculations in both the continuum basis, with a large momentum-space cutoff, and the band-projected basis, where the number of included bands is treated as a tunable parameter.
The details of the calculation is presented in the Supplemental Material (SM) \cite{SM}. 

At filling factor $\nu=1$, our self-consistent mean-field calculation reveals a strong tendency toward
 spontaneous spin-valley ordering, leads to various insulating ground states. 
 At $D=0$, the ground state is a ferromagnetic (i.e. spin-valley polarized) quantum anomalous Hall insulator (QAHI) 
 with Chern number $C=\pm1$, corresponding to spin up or spin down  polarization.
 Figure~\ref{MF_band}(a) plots the mean-field band structure at $D=0$ for the case of a spin-up polarization.
 As $D$ increases, the ground state undergoes a topological transition 
 from the QAHI phase to a trivial FM insulator state with $C=0$.
 Figure~\ref{MF_band}(b) shows the ground state phase diagram as a function of displacement field at twist angle $\theta=3.5^{\circ}$, where the blue (red) line plots the charge (spin) gap, which is the energy gap for an independent charge (spin-flip) excitation.
The phase diagram also exhibits a region at large displacement field where the FM phase becomes unstable toward an intervalley coherent order.
This instability is signaled in our calculation by the magnon energy becoming negative, as discussed in detail below, and is consistent with the onset of an intervalley coherent phase obtained in mean-field calculations including the intervalley channel \cite{Pan2020,WuTMDPhaseDiagram2023,WuNu1Topology2024,MillisMoireHubbard2021,KennesDensityWave2023,MillisIVCAFMSC2024,BernevigFCIMoTe22024,XiaoAFCI2024,Adam2025,Millis2025}.
 Here, we focus on the FM phases that are most relevant experimentally for homobilayers at small to intermediate twist angles.
 
\begin{figure}[t!]
	\centering
	\includegraphics[width=0.475\textwidth]{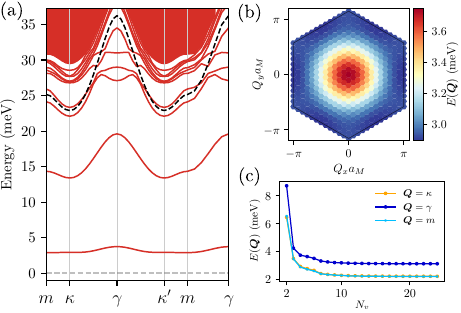}
	\caption{
		\label{magnonband}(a) Magnon bandstructure in the FM state at $D=0$.  The black dashed line marks the minimum energy of the spin-flip inter-band continuum. (b) Color plot of the lowest magnon band energy with band maximum (minimum) at the $\bm{\gamma}$ ($\bm{m}$). (c) Convergence of the magnon energy at high symmetry momentum, $\bm{\kappa},\bm{\gamma},\bm{m}$ as a function of the number of filled spin-down bands, $N_v$, in the TDHF calculation (as illustrated in Fig.~\ref{MF_band}(c)).
	}
\end{figure}

\emph{Collective spin excitations}---\noindent
Spin fluctuations around the FM ground state consist of spin-flip excitations from the filled spin-down bands 
to the topmost empty spin-up band, as illustrated in Fig.~\ref{MF_band}(c). 
These excitations are coupled via the Coulomb interaction, giving rise to 
collective eigenmodes of spin fluctuations, i.e., magnons.
We adopt the time-dependent Hartree-Fock (TDHF) method 
to calculate the magnon spectrum on top of the mean-field ground state\cite{TBGCollectiveMode2021}.
It involves solving the equation of motion of the spin-flip
excitation operator, $\hat{c}^{\dagger}_{1,\uparrow,\bmk'}\hat{c}_{n,\downarrow,\bmk}$, 
in momentum and energy representation.
Here $\hat{c}^{\dagger}_{1,\uparrow,\bmk'}$ is the creation operator for the empty topmost spin-up band, and $\hat{c}_{n,\downarrow,\bmk}$ is the annihilation operator for the filled spin-down bands 
with $n=1,..., N_v$ labeling the band index.

\begin{figure}[t!]
	\centering
	\includegraphics[width=0.475\textwidth]{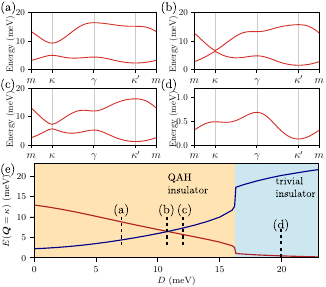}
	\caption{
		\label{topotransition} 
		Isolated low-energy magnon bands at representative displacement fields across the topological transition: (a) $D=7.0$ meV, (b) $D=10.5$ meV, (c) $D=12.0$ meV, and (d) $D=20.0$ meV. (e) Displacement-field dependence of the two lowest magnon energies at $\bm{\kappa}$, where the magnon band gap closes. The vertical dashed lines mark the displacement fields corresponding to panels (a)–(c). 
	}
\end{figure}

By defining the spin-flip excitation basis 
$|n,\bmk,\bmQ\rangle=\hat{c}^{\dagger}_{1,\uparrow,\bmk+\bmQ}\hat{c}_{n,\downarrow,\bmk}|\Omega_{FM}\rangle$,
the TDHF equation can be converted into an eigenvalue problem:
\begin{align}
	AX=\omega  X
\end{align}
where 
$A^{nn'}_{\bmk,\bmk';\bmQ}=-\langle n,\downarrow,\bmk;1,\uparrow,\bmk'+\bmQ |V| n',\downarrow,\bmk';1,\uparrow,\bmk+\bmQ\rangle$ is Coulomb matrix elements in the exchange channel
and $X_{n,\bmk,\bmQ}$ is a vector in the $|n,\bmk,\bmQ\rangle$ basis. 
We set $\hbar=1$ throughout this work.
$\bmQ$ is the center-of-mass mass momentum and different momentum  sectors do not couple with each other due to momentum conservation, 
so $A$ is diagonal in $\bmQ$.
(See more details in the SM\cite{SM})
Diagonalizing the eigenvalue problem leads to the magnon eigenstate
\begin{align}
	|\Psi_{m,\bmQ}\rangle = \sum_{n,\bmk}Z^m_{n,\bmk,\bmQ}|n,\bmk,\bmQ\rangle
\end{align}
with energy $\omega_{m,\bmQ}$ ,
where $Z^m_{n,\bmk,\bmQ}$ is the complex coefficient and $m=1,2,\dots,N_{\bmk}N_{v}$ the magnon band index.
$N_{\bmk}$ is the number of momentum points in moir\'e Brillouin zone (mBZ)
and $N_v$ is the number of spin down bands included in the calculation.

The magnon band structure at $D=0$ features two isolated low-energy bands
that emerge as collective modes of Coulomb-coupled spin-flip excitations, as shown in Fig.~\ref{magnonband}(a). 
The two magnon bands lie well below the interband spin-flip excitation continuum, marked by the black dashed line.
Fig.~\ref{magnonband}(b) plots the colormap of the energy dispersion of the lowest magnon band over the mBZ,
where the magnon gap, defined as the minimum magnon energy, $\Delta_{\rm mag}=2.3$ meV, 
is located at finite momentum $\bmQ=\bm{m}$.
The two magnon bands carry Chern numbers $C=\pm 1$, 
with the signs reversed for the opposite valley-polarized ferromagnetic state
\cite{WuMagnonsMoTe22025, ZhouMagnonsNSR2026}.
Our numerical results reveal that the convergence of the magnon energy
depends on the number of filled spin-down bands $N_v$ included in the calculation. 
Fig.~\ref{magnonband} (c) shows the magnon energy at high-symmetry momentum as a function of $N_v$
for the un-projected calculation 
(see SM~\cite{SM} for results from the band-projected case),
which shows that at least $N_v \geq 4$ is required to reach near convergence.

\emph{Field-induced topological transition}---\noindent
The presence of two isolated low-energy magnon bands at $D = 0$ is not accidental, 
but hints at a nontrivial internal structure of the magnon excitation.
To understand this, we examine the evolution of the magnon spectrum with displacement field. Figure~\ref{topotransition}(a)–(d) shows the low-energy magnon bands at displacement fields 
$D = 7.0$~meV, $10.5$~meV, $12.0$~meV, and $20.0$~meV, respectively.
As $D$ increases, the gap between the two magnon bands decreases and closes
at $D^*=10.5$ meV, where the bands undergo a topological quantum phase transition.
The Chern number of the lowest magnon band changes from $C=1$ to $C=0$.
The gap closing occurs at $\bmQ = \bm{\kappa}$ ($\bm{\kappa}'$) for positive (negative) displacement field.
Notably, this magnon topological transition does not coincide with the ground-state topological transition, 
which occurs at a higher displacement field. Instead, it takes place within the QAHI phase, 
where the ground-state Chern number remains unchanged, as shown in Fig.~\ref{topotransition}(e). 
At the QAHI–trivial insulator transition, the magnon bands remain topologically trivial. 
As $D$ increases beyond the transition, the second magnon band shifts to higher energy 
and merges into the interband spin-flip continuum, leaving only a single low-energy magnon band.

For spins localized on a lattice, magnons correspond to collective precession 
of spins around the ordered moment direction. 
In TMD moiré homobilayers, the low-energy electron bands can be mapped 
onto an effective Kane-Mele model on a honeycomb lattice. 
While a ferromagnet on a honeycomb spin lattice supports two magnon bands, 
it consists of two localized spins per unit cell, i.e., filling factor $\nu = 2$. 
However, for filling factor $\nu=1$ considered here, having two magnon bands implies 
that the magnetization within each moiré unit cell cannot be described by a single average spin. 
Instead, the two magnon branches emerge as different excitation modes 
with orthogonal internal spin textures within each unit cell. 
At $D = 0$, the electron density in each moiré unit cell is equally distributed between top and bottom layers, 
with maxima localized near the MX and XM high-symmetry sites. 
The two magnon bands correspond to collective precession of the partial spin magnetizations on the two layers, 
either in phase (symmetric mode) or out of phase (antisymmetric mode). 
As $D$ increases, the electrons become polarized in one layer, and the system approaches a triangular spin lattice configuration supporting only a single, topologically trivial magnon band.

\begin{figure}[t!]
	\centering
	\includegraphics[width=0.475\textwidth]{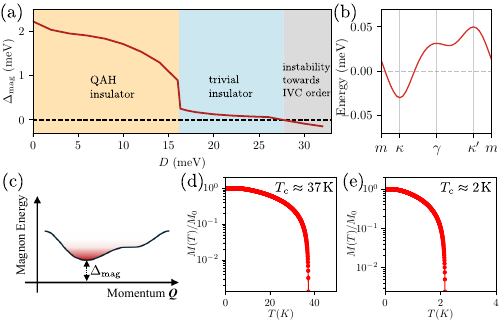}
	\caption{
		\label{mgap}(a) Magnon gap ($\Delta_{mag}$) as a function of displacement field across different FM phases indicated by the shaded color. At $D>27.2$ meV, an instability develops toward an intervalley coherent density wave order (gray shaded region). (b) The lowest magnon band at $D=28$ meV showing the instability at finite momentum $\bm{Q}=\bm{\kappa}$. (c) Schematic illustration of magnon thermal population at finite temperature. 
		Temperature dependence of the magnetization in (d) the QAH  phase and (e) trivial insulator phase.
	}
\end{figure}

\emph{Magnon gap and transition temperature}---\noindent
The presence of a finite magnon gap is essential for stabilizing long-range magnetic order 
at finite temperature in two-dimensional systems, 
where proliferation of thermal excitations of gapless magnon modes would otherwise destabilize the ordered phase.
The magnon gap arises as a consequence of magnetic anisotropy.
In monolayer TMDs, the spin-valley locked low-energy states 
$|\negmedspace+\negmedspace K,\uparrow\rangle$ and $|\negmedspace-\negmedspace K,\downarrow\rangle$ 
define a pseudospin doublet within an approximate SU(2) symmetry under simultaneous rotation of spin and valley. Magnetic anisotropy is therefore almost absent, and indeed long-range magnetic order is not observed in pristine monolayers.
For TMD moir\'e superlattices, the interlayer coupling $T_s(\bmr)$ explicitly breaks this symmetry,
enabling the emergence of a finite magnon gap.

In the ferromagnetic phase, the magnon gap depends on the interlayer displacement field $D$, as shown in Fig.~\ref{mgap}(a), which controls the layer polarization and interaction-induced interlayer coherence.
Overall, the magnon gap decreases as $D$ increases, reflecting the suppressed effect of the symmetry-breaking interlayer hybridization.
Most notably, the magnon gap drops drastically across the transition from the QAHI to the trivial ferromagnetic phase, decreasing by nearly an order of magnitude—from $\sim\!1$–2 meV in the QAHI phase to $\sim\!0.2$ meV in the trivial phase.
This suggests a qualitative difference in the stability of magnetic order due to the difference in the
ground state topology.
It is consistent with the microscopic picture that the nontrivial band topology originates from the interlayer hybridization which in turn gives rise to the magnetic anisotropy.
In the trivial phase, the interlayer hybridization is suppressed due to the strong layer polarization,
thereby suppressing the magnetic anisotropy.

As $D$ varies, the magnon minimum shifts in momentum space, so a low-energy expansion in terms of a gap and spin stiffness around the minimum depends sensitively on its location. Both quantities admit a geometric interpretation in terms of the underlying electron bands \cite{WuDasSarma2020, Yu2025npjQM, Rossi2021COSSMS, CanoQuantumDipole2025,  Arita2026}. 
Here, we remark that the relevance of such an expansion is determined in the first place by the overall magnon band structure, which is itself sensitive to microscopic details. This is particularly evident across the topological transition, beyond which the lowest magnon band becomes very flat in the trivial FM phase, with a bandwidth below $0.5$ meV.

Upon further increasing $D$, the magnon gap becomes negative near $D = 28$ meV, 
signaling an instability of the ferromagnetic phase. 
As shown in Fig.~\ref{mgap}(b), the softening occurs at finite momentum $\bmQ = \bm{\kappa}$, 
suggesting a transition to a translational symmetry-breaking spin-density wave, 
likely an intervalley coherent (IVC) state, as typically found in Hartree-Fock calculations at large displacement field.

At finite temperature, thermal occupation of magnon bands reduces the spin magnetization, as schematically illustrated in Fig.~\ref{mgap}(c).
Since each magnon corresponds to a reduction by one $\mu_B$ per moir\'e unit cell relative to the ferromagnetic ground state, the spin magnetization $M(T)$ at finite temperature is given by
\begin{align}
	\frac{M(T)}{M_0} = 1 - \frac{1}{N_{\bmQ}}\sum_{m, \bmQ} n_{B}(E_{m,\bmQ}, T)
\end{align}
where $n_B(E_{m,\bmQ}, T)$ is the Bose–Einstein distribution for magnons with energy $E_{m,\bmQ}$ and
$m$ the magnon band index. 
$M_0=\mu_{B}/A_{\text{moir\'e}}$ is the zero-temperature spin magnetization 
where $\mu_B$ is the Bohr magneton and $A_{\text{moir\'e}}$ is the moir\'e unit cell area.
$N_{\bmQ}$ is the number of magnon momentum which equals to $N_{\bmk}$.
Figure~\ref{mgap}(c) and (d) shows the temperature dependence of the magnetization for states
in the QAHI phase ($D=0$ meV) and trivial phase ($D=20$ meV), respectively.
The FM Curie temperature $T_c$ can be determined by the temperature
at which the magnetization approaches zero, $M(T_C)=0$.
In the QAHI phase [Fig.~\ref{mgap}(c)], we find $T_c = 37$ K. 
By contrast, in the trivial phase [Fig.~\ref{mgap}(d)], the Curie temperature 
drops to $T_c = 2$ K, more than an order of magnitude smaller.
Experimentally, the Curie temperature measured in the $D = 0$ meV QAHI phase is approximately 14 K, 
whereas clear signatures of ferromagnetic ordering in the trivial phase remain elusive. 
This observation is consistent with our finding of a significantly suppressed $T_c$ in the trivial regime.
We note that our approach for obtaining $T_c$ ignores magnon-magnon interactions,
which renormalize the magnon band energy and can lead to an overestimation of the transition temperature.

\emph{Discussions}---\noindent
We have investigated the nature of collective spin excitations 
and their influence on the stability of ferromagnetic order 
in twisted TMD homobilayers at filling factor $\nu = -1$. 
In the QAHI phase, the nontrivial topology of the ground state 
endows the magnon excitations with a topological character 
that undergoes a transition as the interlayer displacement field is varied. 
The two isolated low-energy magnon bands observed in this regime
cannot be described by conventional spin-lattice models and 
therefore do not correspond to simple spin precession modes. 
Instead, they reflect a nontrivial internal spin texture distributed 
between two high-symmetry sites within the moiré unit cell. 
In contrast, the trivial ferromagnetic phase hosts a single low-energy magnon band 
consistent with spin precession on a triangular lattice. 
Most notably, the magnon gap exhibits a pronounced suppression 
across the QAHI-to-trivial insulator transition. This behavior is 
attributed to the magnetic anisotropy originating from spin-dependent interlayer hybridization 
and is continuously tunable by the displacement field.
The ferromagnetic Curie temperature is significantly enhanced in the QAHI phase due to the combined effects of interlayer hybridization and interaction-induced interlayer coherence. In the trivial ferromagnetic phase, the Curie temperature is more than one order of magnitude smaller, in agreement with the absence of observed trivial ferromagnetism in twisted MoTe$_2$ at large displacement fields, which has hitherto remained unexplained.

Our findings highlight the potential of twisted TMD moiré homobilayers as a versatile platform for exploring tunable magnetism in two dimensions. Recent experimental progress on propagating neutral modes and optical control or switching of magnetism in TMD moir\'e systems further underscores the importance of understanding collective magnetic excitations in these materials \cite{JinNeutralModes2025,ImamogluOpticalChern2025,XuOpticalFCI2025,GaoOpticalSwitching2025}. More broadly, the ability to control the magnetic anisotropy and Curie temperature through an external electric field suggests a promising route for engineering high-temperature two-dimensional magnets, even in materials that are intrinsically nonmagnetic.

\begin{acknowledgments}
	{\em Acknowledgments}---\noindent	
	The authors acknowledge helpful discussions with Chenhao Jin and Allan H. MacDonald. 
	This work was supported by the Laboratory for Physical Sciences through its support of the Condensed Matter Theory Center.
\end{acknowledgments}

\end{document}